\documentclass[a4paper,fleqn,usenatbib]{mnras}

\usepackage[T1]{fontenc}
\usepackage{ae,aecompl}

\usepackage{graphicx}	
\usepackage{amsmath}	
\usepackage{amssymb}	

\title[Extended stellar substructure surrounding the Bo\"{o}tes\,I dwarf spheroidal galaxy]{Extended stellar substructure surrounding the Bo\"{o}tes\,I dwarf spheroidal galaxy}

\author[T. A. Roderick et al.]{
T. A. Roderick,$^{1}$\thanks{E-mail: Tammy.Roderick@anu.edu.au}
A. D. Mackey,$^{1}$
H. Jerjen,$^{1}$
G. S. Da Costa$^{1}$
\\
$^{1}$Research School of Astronomy and Astrophysics, Australian National University, Canberra, ACT 2611, Australia\\
}

\date{Accepted XXX. Received YYY; in original form ZZZ}

\pubyear{2015}

\begin{document}
\label{firstpage}
\pagerange{\pageref{firstpage}--\pageref{lastpage}}
\maketitle

\begin{abstract}
We present deep stellar photometry of the Bo\"{o}tes\,I dwarf spheroidal galaxy in $g$ and $i$ band filters, taken with the Dark Energy Camera at Cerro Tololo in Chile. Our analysis reveals a large, extended region of stellar substructure surrounding the dwarf, as well as a distinct over-density encroaching on its tidal radius. A radial profile of the Bo\"{o}tes\,I stellar distribution shows a break radius indicating the presence of extra-tidal stars. These observations strongly suggest that Bo\"{o}tes\,I is experiencing tidal disruption, although not as extreme as that exhibited by the Hercules dwarf spheroidal. Combined with revised velocity dispersion measurements from the literature, we see evidence suggesting the need to review previous theoretical models of the Bo\"{o}tes\,I dwarf spheroidal galaxy.
\end{abstract}




\section{Introduction}
\label{sec:intro}
The study of the origin and nature of the Milky Way (MW) dwarf galaxy population has received increased attention since the advent of digital all-sky surveys. As a result, a new class of ultra-faint stellar systems has been discovered \citep[e.g.][]{2007ApJ...654..897B,2007ApJ...662L..83W,2007ApJ...669..337K,2015ApJ...807...50B,2015ApJ...805..130K,2015ApJ...799...73K,2015ApJ...803...63K,2015ApJ...813...44L,2015ApJ...813..109D}. According to the concordance cosmological model, $\Lambda$CDM, the dwarf galaxy population is one of the visible remnants of the galaxy formation process. In this context, galaxies have formed through the accretion and merger of baryonic matter embedded in dark matter halos \citep[see:][]{1999ApJ...524L..19M,2007ApJ...667..859D,2011ASL.....4..297D}. This scenario is supported by the high dark matter content inferred in the dwarf galaxies \citep[e.g.][]{2007ApJ...670..313S,2007MNRAS.380..281M,2011ApJ...733...46S,2015ApJ...810...56K,2015ApJ...811...62K}. However, observations show that most of the currently known MW satellite galaxies are aligned in a vast polar structure \citep{2015MNRAS.453.1047P}, said to be inconsistent with a primordial origin. A similar structure seen amongst the Andromeda satellites \citep{2013Natur.493...62I} has lent support to an alternative explanation where the MW satellite population is largely made up of tidal remnants from the interaction of the MW with another galaxy \citep[see:][]{Kroupa:2005iy,2013MNRAS.435.1928P,2014MNRAS.442.2419Y,2014ApJ...789L..24P}.

In order to resolve the conflict between these fundamentally different scenarios and gain a better understanding of the MW neighbourhood, further investigation is required. In depth studies of the MW dwarf galaxy population have demonstrated that a number of these dwarfs show signs of tidal disruption \citep[e.g.][]{1994Natur.370..194I,2003AJ....125.1352P,Battaglia:2012it,2013AJ....145..163N,2014MNRAS.444.3139M,Roderick:2015jj}.  A result of this is that velocity dispersion measurements of such galaxies can be inflated by unbound stars \citep{2007MNRAS.378..353K}.  Moreover, mass-to-light ratios determined from kinematic samples assume a system is in dynamic equilibrium.  Where this is not the case, the mass-to-light ratio can be over-estimated. Furthermore, several cases exist in which improved membership determination has led to considerably lower velocity dispersion estimates, and consequently mass-to-light ratios \citep{2009MNRAS.394L.102L,Koposov:2011kx}.

Our recent work \citep{Roderick:2015jj,submitted} has demonstrated the effectiveness of deep, wide-field, photometric observations in revealing extended stellar substructure associated with MW dwarf galaxies, leading to improved understanding of their current dynamic state.  Following on from this, we present here a wide-field, photometric study of the Bo\"{o}tes\,I dwarf spheroidal galaxy.

The Bo\"{o}tes\,I dwarf (see Table \ref{tab:parameters}), discovered in the Sloan Digital Sky Survey \citep[SDSS, ][]{2006ApJ...647L.111B}, is relatively unstudied compared to many of its other MW satellite companions. Bo\"{o}tes\,I is a low luminosity \citep[$M_V = -5.92\pm0.2$,][]{2012ApJ...744...96O}, metal poor \citep[Fe/H = $-2.55 \pm 0.11$,][]{2010ApJ...723.1632N} system, and is one of the most gas poor dwarfs known \citep{2007MNRAS.375L..41B}. It shows an elongated stellar distribution \citep{2006ApJ...647L.111B,2012ApJ...744...96O}, and has a high mass-to-light ratio \citep[$>100$,][]{2006ApJ...650L..51M,2007MNRAS.380..281M,Koposov:2011kx}. Results from N-body simulations, using a number of models of Bo\"{o}tes\,I, have led to the conclusion that its progenitor must have been dark matter dominated, since a purely baryonic star cluster could not reproduce the observed velocity dispersion \citep{Fellhauer:2008dt}.  However, more recent work by \citet{Koposov:2011kx} has demonstrated that Bo\"{o}tes\,I not only has a much lower velocity dispersion than previously thought, it also exhibits a secondary hot component alongside its dominant, colder component. This calls into question our previous understanding of the dark matter content of Bo\"{o}tes\,I, and presents a potentially interesting dynamic history.  

Study of the stellar population of Bo\"{o}tes\,I has shown it to be old \citep[$> 13$Gyr,][]{2012ApJ...744...96O}, and consistent with a single epoch, short period burst of star formation \citep{2006ApJ...647L.111B,2012ApJ...744...96O}. Bo\"{o}tes\,I has a population of variable stars \citep{2006ApJ...649L..83S,2006ApJ...653L.109D}, and blue straggler stars (BSS) \citep{2013ApJ...774..106S}, and shows chemical properties consistent with essentially primordial initial abundances \citep{Gilmore:2013gy}.  Despite its projected proximity to the Sagittarius stream, any association has been ruled out \citep{2010ApJ...718.1128L}.

Although its elongation and irregular shape were noted upon its discovery, as well as more recently \citep{2006ApJ...647L.111B,2012ApJ...744...96O}, little has been done to characterise the spatial extent of Bo\"{o}tes\,I quantitatively.  This may in part be due to the difficulty in observing such a faint object, particularly on any broad spatial scale.  However, the emergence of wide-field CCD cameras such as the Dark Energy Camera \citep[DECam,][]{2015AJ....150..150F}, on the 4m Blanco telescope at Cerro Tololo in Chile, makes such observations possible.  Here we present deep, wide-field observations of the Bo\"{o}tes\,I dwarf spheroidal galaxy with the intention of conducting a quantitative analysis of stellar substructure associated with the system.  In Section \ref{sec:obs} we present our observations and data reduction.  This is followed by foreground discrimination and spatial mapping in Sections \ref{sec:fgd} and \ref{sec:mapping}. We perform a brief analysis of the structural parameters of Bo\"{o}tes\,I in Section \ref{sec:parameters}. In Section \ref{sec:analysis} we present our substructure analysis and methodology.  Finally, Sections \ref{sec:disc} and  \ref{sec:sum} present our discussion and summary.

\begin{table}
\begin{center}
\caption{Fundamental Parameters of Bo\"{o}tes\,I}
\begin{tabular}{llc}
\hline\hline
Parameter & Value & Ref.$^a$\\
\hline
RA (J2000) & 14:00:06 & 1 \\
DEC (J2000) & $+$14:30:00 & 1 \\
$l$ & $358.1^\circ$ & 1\\
$b$ & $69.6^\circ$ & 1\\
$D_\odot$ & $65\pm3$\,kpc & 2 \\
$(m-M)_0$ & $19.07 \pm 0.11$ & 2 \\
$r_h$ & $12.5\arcmin\pm0.3\arcmin$ & 2 \\
$M_V$ & $-5.92\pm0.2$\,mag & 2\\
$\sigma$ & $2.4^{+0.9}_{-0.5}$ km s$^{-1}$ & 3\\
$v_\odot$ & $101.8 \pm 0.7$ km s$^{-1}$ & 3\\
{\rm [Fe/H]}& $-2.55 \pm 0.11$\,dex & 4\\
$\theta$ & $13.8^\circ \pm 3.9^\circ$ & 5\\
$\epsilon$ & $0.26 \pm 0.01$ & 5\\
$r_t$ & $31.6\arcmin\pm4.0\arcmin$ & 5\\
\hline\hline
\label{tab:parameters}
\end{tabular}
\end{center}
\textbf{$^a$References:} (1) \citet{2006ApJ...647L.111B}, (2) \citet{2012ApJ...744...96O}, (3) \citet{Koposov:2011kx}, (4) \citet{2010ApJ...723.1632N}, (5) This work
\end{table}

\section{Observations and Data Reduction}
\label{sec:obs}
Observations were carried out as part of observing proposal 2013A-0617 (PI: D. Mackey), on 2013 February 15, using DECam on the 4m Blanco telescope at Cerro Tololo in Chile. DECam is comprised of a hexagonal mosaic of 62 2K$\times$4K CCDs, each with a pixel scale of $0\farcs27$/pix, creating a total field-of-view of 3 square degrees. 

\begin{figure}
\centering
\includegraphics[width=\columnwidth]{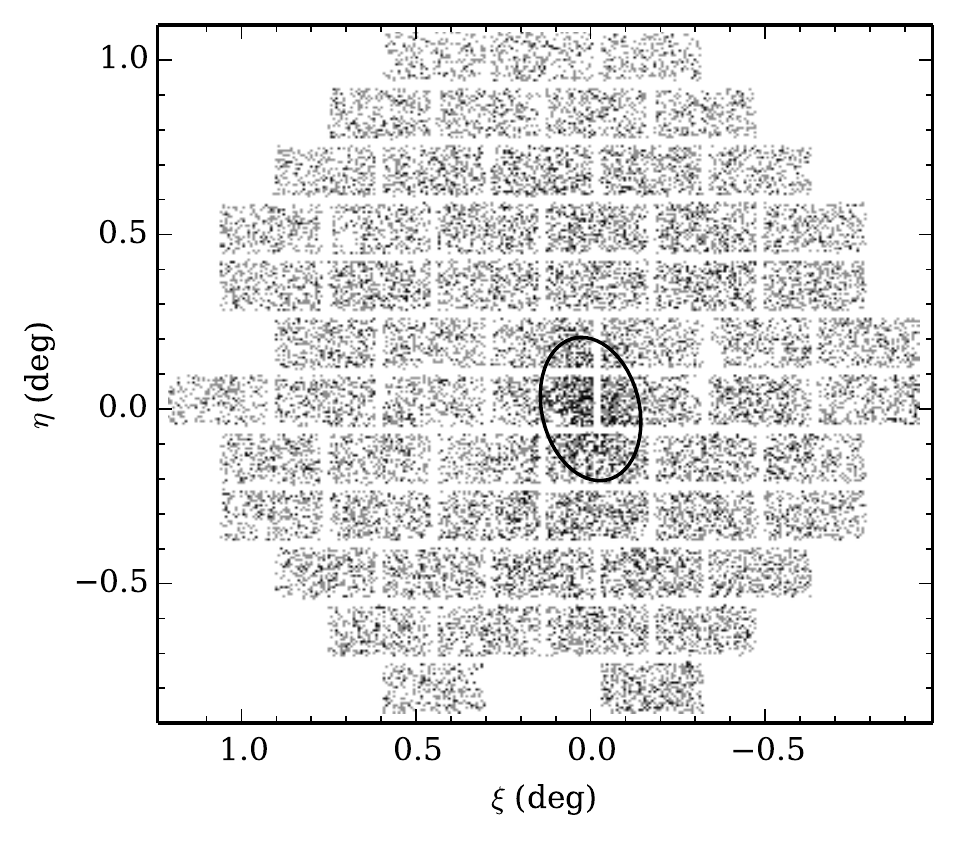}
\caption{All point sources in our DECam field-of-view. North is up, and East is to the left. The black ellipse denotes the half-light radius of Bo\"{o}tes\,I. Note that two CCDs were faulty at the time of observing, and have been masked out.  There are two bright stars in close proximity to one another on the western side of the field, leading to a lack of detections in that area. This is discussed later in the paper.}
\label{fig:field}
\end{figure}

The data set consists of a single pointing taken in the direction of Bo\"{o}tes\,I (see Figure \ref{fig:field}), in $g$ and $i$ band filters.  A total of $3\times300$s exposures were taken in each filter, providing a total integration time of 900s each, with an average seeing of  $1\farcs30$ and  $1\farcs14$ in $g$ and $i$ respectively.

The raw images were processed with the DECam community pipeline\footnote{http://www.ctio.noao.edu/noao/content/Dark-Energy-Camera-DECam}  \citep{2014ASPC..485..379V}.  This processing included the subtraction of sky-background, WCS solution fitting, and the co-addition of images into a single, deep, multi-extension FITS image for each filter. Each of these images is presented as nine image slices, or tiles, contained in nine separate extensions of the FITS file.

\subsection{Photometry and Star/Galaxy Separation}
The photometry was carried out using the pipeline described by \citet{Roderick:2015jj}.  This involved several key steps outlined in the following.  

First, WeightWatcher\footnote{http://www.astromatic.net/software/weightwatcher} \citep{2008ASPC..394..619M} was used in combination with the weight maps from the DECam community pipeline, in order to mask out non-science pixels in the images prior to the performance of aperture photometry.  The photometry was then conducted with Source Extractor\footnote{http://www.astromatic.net/software/sextractor} \citep{1996A&AS..117..393B} in two passes on each image.  The first pass was shallow, providing an estimate of the average FWHM ($\overline{F}$) across the field.  A second, deeper pass was then performed, using this information as input for aperture sizes.  Flux was measured inside $1\times \overline{F}$ and  $2\times \overline{F}$, separately on each of the nine image tiles for each filter. This allowed the aperture size to vary across the large field-of-view, allowing for fluctuations in the point-spread-function.

\begin{figure}
\centering
\includegraphics[width=\columnwidth]{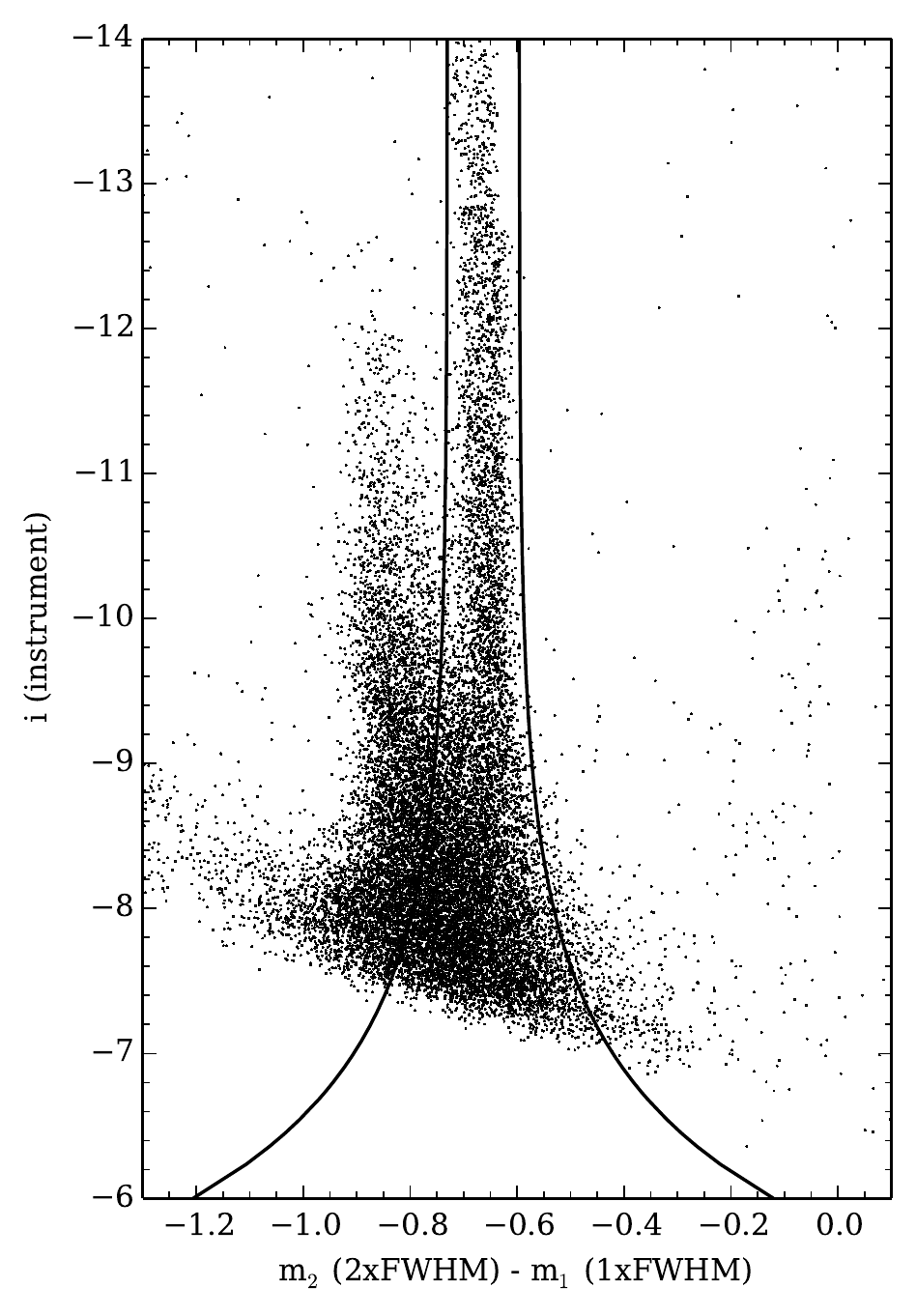}
\caption{Magnitude difference between apertures for each source extracted in the central $i$-band image tile. Stellar objects are shown as a clear locus, with a selection region shown by black lines. The selection region was determined by fitting an exponential to the distribution of stars and reflecting it about the locus.}
\label{fig:stargal}
\end{figure}

Once the photometry was complete, star/galaxy separation was performed.  The technique used is described in detail in \citet{Roderick:2015jj}, and involves taking the magnitude difference between apertures for each source in the catalogue, revealing a clear stellar locus in the data. Figure \ref{fig:stargal} demonstrates this process for the central image tile of Bo\"{o}tes\,I. and clearly shows the stellar locus. The locus is isolated by fitting an exponential to the distribution to the right of the mean, and reflecting it to create a selection region. This is combined with the Source Extractor classification flag (flag > 0.35) and achieves star/galaxy separation to a depth over a magnitude greater than that achieved by Source Extractor alone. As seen in \citet{submitted}, the Source Extractor classification flag can be inconsistent when used across multiple fields with large variations in seeing. However, for individual fields (such as Bo\"{o}tes\,I), when used conservatively it can assist in star/galaxy separation for faint objects \citep{Roderick:2015jj}. The star/galaxy separation was performed on the $i$-band catalogue prior to cross-matching with the $g$-band catalogue.  The product of this process was a single catalogue of stellar sources, containing the photometry for both filters. 

After completion of the catalogue, it was calibrated to the SDSS photometric system.  For that purpose, the catalogue was cross-matched with stars from SDSS data release 10 \citep[SDSS DR10:][]{2014ApJS..211...17A}.  The SDSS magnitudes were compared to the catalogue magnitude inside the $2\times \overline{F}$ aperture, and a zero point ($ZP$) and colour term ($c$) determined for each filter, where colour is defined as $(g-i)_{inst}$. These values, determined by a least squares fit, are summarised in Table \ref{tab:data}.  Once calibrated, a correction for Galactic extinction was applied.  This has a greater effect on the $g$-band than the $i$-band, with $0.048<A_g<0.107$ (approximately $3\%$ differential variation in magnitude across the field).  Rather than applying a single correction based on the average extinction, the correction factor was allowed to vary across the field.  This was achieved in a simple manner by making use of the calibration to SDSS DR10. Each star in the catalogue was matched to its nearest SDSS neighbour, and the corresponding correction value used.  According to \citet{2002AJ....123..485S}, the extinction correction values in SDSS DR10 are given by $A_g=3.793E(B-V)$ and $A_i=2.086E(B-V)$, and are based on the extinction maps of \citet{1998ApJ...500..525S}.

\begin{table}
\begin{center}
\caption{Properties of the observations taken for Bo\"{o}tes\,I.}
\begin{tabular}{lcc}
\hline\hline
 Property & $g$ & $i$\\
\hline
Average seeing & $1\farcs30$ & $1\farcs14$\\
$ZP$ & 31.124 & 31.314\\
$c$ & 0.069 & 0.078\\
50\% Completeness & 24.69 & 23.94\\
\hline\hline
\label{tab:data}
\end{tabular}
\end{center}
\end{table}

\subsection{Completeness}
\begin{figure}
\centering
\includegraphics[width=\columnwidth]{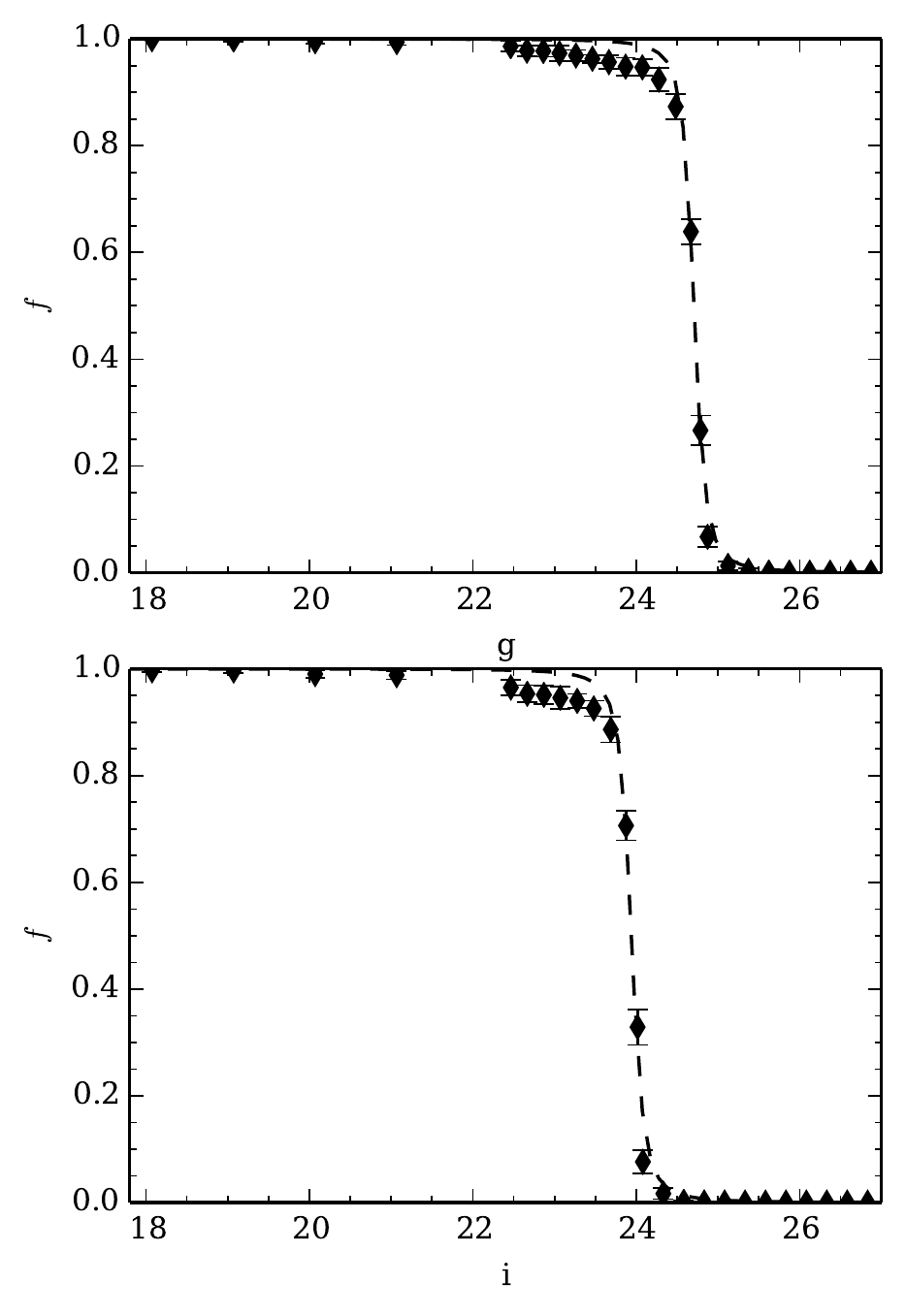}
\caption{Top and bottom panels show the results of the completeness test in $g$ and $i$ respectively. The fitting relation described by \citet{1995AJ....109.1044F} is shown by the dashed lines. The 50\% completeness limits determined by the fitting relation are given in Table \ref{tab:data}.}
\label{fig:completeness}
\end{figure}

A completeness test was performed in order to assess the limiting magnitudes and photometric uncertainties in each filter. This was achieved by adding artificial stars to each image with the IRAF \textit{addstar} task.  Artificial stars were added to each image in magnitudes ranging between 18.0 and 26.75 mag, with randomly assigned coordinates.  Each image was then processed through the photometry pipeline to recover the artificial stars.  This process was repeated 30 times, with a total of 145,800 artificial stars added to each image. The resulting completeness curve for each filter is shown in Figure \ref{fig:completeness}.  The relation from \citet{1995AJ....109.1044F} was fit to each completeness curve in order to determine the 50\% completeness limit.  These relations are shown in Figure \ref{fig:completeness}, and the limiting magnitudes listed in Table \ref{tab:data}.  The catalogue was cut at the 50\% limiting magnitude in each filter and is limited by the $i$-band observations, with the 50\% limiting magnitude 0.75 mag brighter than the $g$-band.  However, the stellar catalogue still reaches down to $i$=23.9\,mag, or approximately 1.5 mag below the main sequence turn off. 

The photometric uncertainties in each band were determined by comparing the recovered magnitude of each artificial star to its input magnitude.  The resulting distributions were each fit with an exponential function, which was then used as a reference to calculate the photometric uncertainty of each individual star in each filter.

\section{Foreground Discrimination}
\label{sec:fgd}
In order to investigate stellar over-densities associated with Bo\"{o}tes\,I, it is essential to first remove as much contamination from MW halo stars as possible.  This is difficult to do, however reasonable discrimination can be achieved with the assistance of a colour-magnitude diagram (CMD) \citep[see][]{Roderick:2015jj,submitted}.  We use the same method as \citet{Roderick:2015jj} to weight stars according to their position on the CMD and proximity to the Bo\"{o}tes\,I fiducial.  The Bo\"{o}tes\,I fiducial is closely approximated using a model isochrone.  This isochrone is then used to create a weight mask by assigning a Gaussian profile in the colour-space, along discrete magnitude intervals.  The Gaussian profile is determined for each magnitude interval using the colour of the isochrone at that magnitude as the peak of the Gaussian, and the photometric uncertainty in colour as the width.  Each Gaussian has a peak value of 1, such that stars closest to the isochrone are given the highest weight.

For the Bo\"{o}tes\,I mask, we selected a model isochrone from the Dartmouth Stellar Evolution Database \citep{2008ApJS..178...89D}, with an age, metalicity and alpha abundance of 15 Gyr, $-2.49$\,dex and $0.2$\,dex, respectively. These values are consistent with the literature, and the isochrone provides an excellent approximation of the Bo\"{o}tes\,I fiducial. Once the mask was created, each star in the catalogue was assigned a weight $w$ between 0 and 1 based on the mask, providing a flexible means of selecting stars consistent with membership of Bo\"{o}tes\,I.  Choosing a larger or smaller weight decreases or increases the size of the selection box around the Bo\"{o}tes\,I stellar population, whilst simultaneously considering the photometric uncertainties in the data.

\begin{figure*}
\centering
\includegraphics[width=\textwidth]{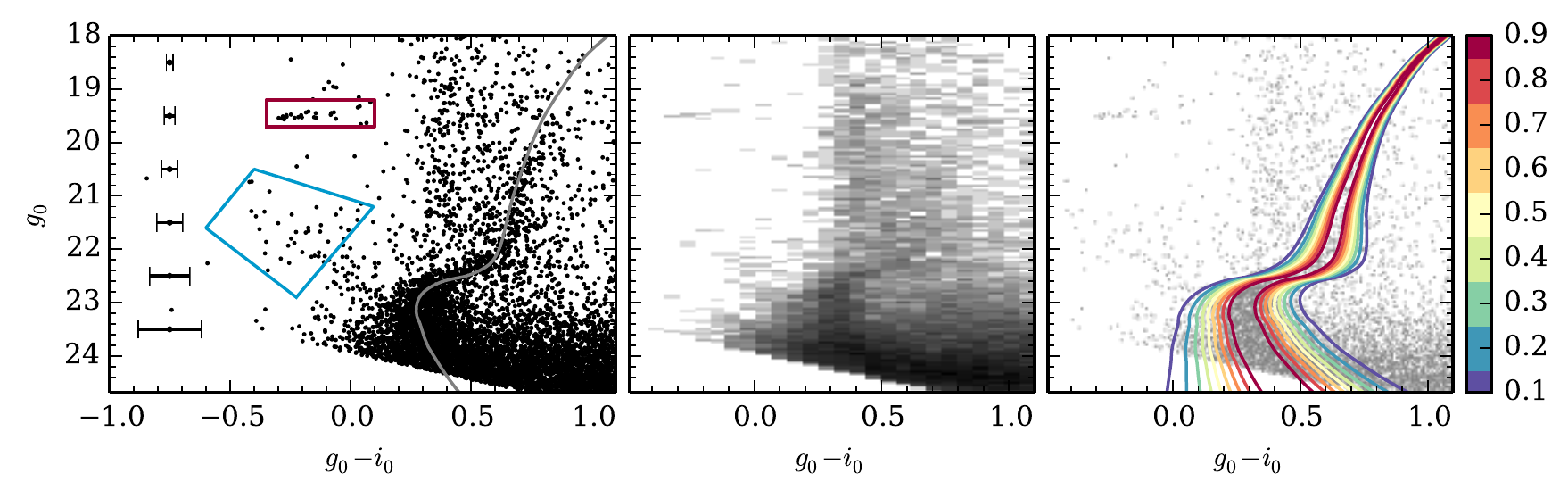}
\caption{Left: Colour magnitude diagram of all stars within the central half-light radius of the Bo\"{o}tes\,I dwarf galaxy.  The red and blue boxes denote the regions of candidate blue horizontal branch and blue straggler stars respectively, which are excluded from the mask. Centre: 2D histogram of the complete Bo\"{o}tes\,I catalogue, showing considerable contamination from MW halo stars. Right: Weight mask generated from the Dartmouth model isochrone that best resembles the Bo\"{o}tes\,I stellar population.  The width of the mask increases with increasing photometric uncertainty at faint magnitudes.}
\label{fig:mask}
\end{figure*}

Figure \ref{fig:mask} illustrates both the Bo\"{o}tes\,I stellar population, and the mask.  The left panel shows the CMD of stars inside the half-light radius \citep[$r_h=12.5\arcmin\pm0.3\arcmin$][]{2012ApJ...744...96O} of Bo\"{o}tes\,I, demonstrating a clear fiducial sequence belonging to the galaxy.  Red and blue boxes also outline candidate blue horizontal branch (BHB) stars and BSS, respectively. They are not included in the mask, but will be used later as a consistency check. The central panel in the figure shows a Hess diagram of the entire DECam field (with bin sizes of 0.07 and 0.06 in $g$ and $g-i$ respectively).  Although Bo\"{o}tes\,I is still apparent, the amount of contamination from the MW foreground is considerable and demonstrates the need for discrimination.  The right hand panel in the figure presents the weight mask, where the values range from 0 to 1. The increasing photometric uncertainties broaden the mask at faint magnitudes.

\section{Spatial Mapping}
\label{sec:mapping}
With each star in the catalogue assigned a weight, we created a map of the spatial distribution of Bo\"{o}tes\,I. This facilitates both the analysis of structural parameters of the dwarf as well as the search for stellar substructure associated with it. The process used follows closely the method used by \citet{Roderick:2015jj,submitted}.

First, the catalogue was separated into two groups, one containing stars consistent with Bo\"{o}tes\,I membership, and the other containing everything else.  This was done in order to create a kind of `foreground' map which could be used to remove the foreground contamination.  A weight of $w\geq0.5$ was chosen as the cutoff weight to create these two groups, since it encompasses as much of the Bo\"{o}tes\,I population as possible without introducing much contamination from the MW (see right panel of Figure \ref{fig:mask}).

\begin{figure*}
\centering
\includegraphics[width=\textwidth]{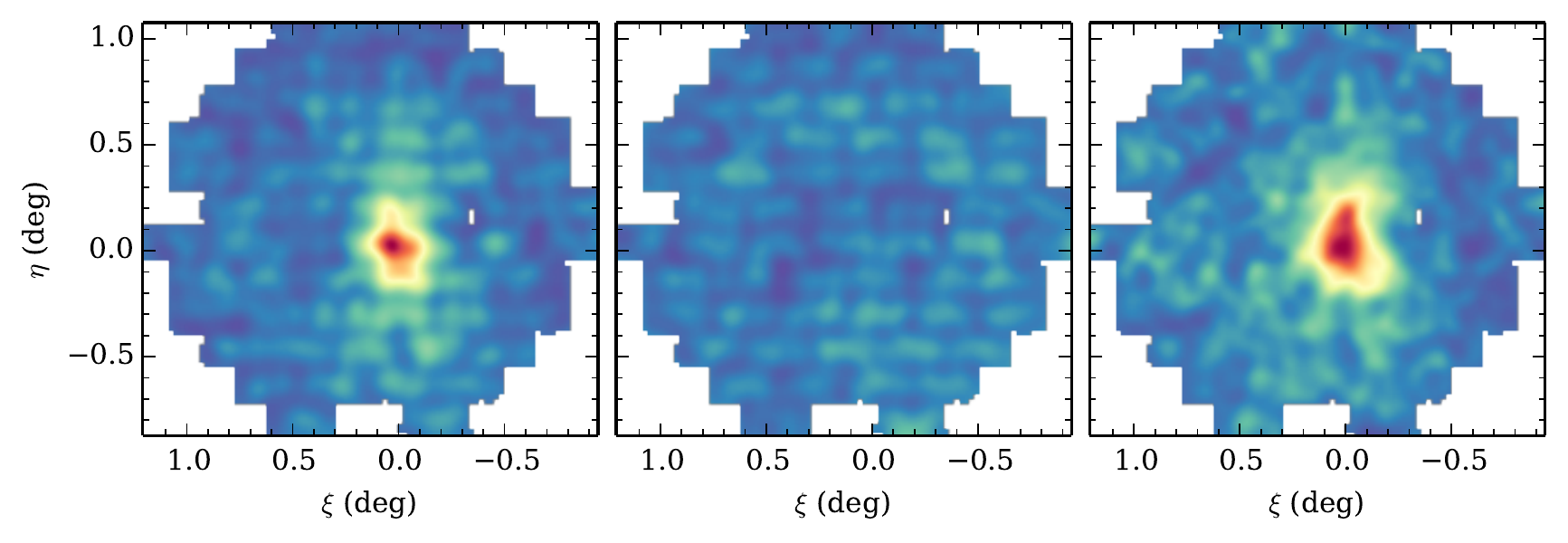}
\caption{The process of creating a smoothed density map for detection of stellar over-densities associated with Bo\"{o}tes\,I.  Left: Smoothed density map of stars consistent with Bo\"{o}tes\,I membership ($w\geq0.5$). Centre: Smoothed density map of stars inconsistent with Bo\"{o}tes\,I membership ($w<0.5$). Right: The smoothed, `foreground' subtracted density map revealing the distribution of stars that are most likely Bo\"{o}tes\,I members. Note that the inter-chip gaps between CCDs in the left and central panels disappeared, highlighting the success of the foreground subtraction. The position of the two bright foreground stars in the field are visible as a white gap in the map. The density scale increases from blue to red in each panel, however while the left and centre panels share the same scale, the right panel has a different scale for visual clarity.}
\label{fig:hist}
\end{figure*}

Once the two groups were defined, they were binned in R.A. and Dec to create normalised (to bin area and sample size) 2D histograms (or pixelated maps).  After experimenting with several different bin sizes, a bin size of $20\arcsec\times20\arcsec$ was chosen since it provided the most detail without too much noise.  The two associated histograms were then smoothed with a Gaussian kernel to create density maps.  Different kernel sizes were trialled, however a smoothing factor of 7 pixels was chosen (equivalent to smoothing over 140\arcsec).  We note that the same major structural features were apparent in the maps independently of the chosen kernel size, however this size again provided the most detail with the least amount of noise.  Once the two density maps were established, the `foreground' map was first used to divide through the Bo\"{o}tes\,I map like a flat field, before being subtracted from the Bo\"{o}tes\,I map to reveal an elongated halo surrounding the Bo\"{o}tes\,I dwarf.  The foreground map was scaled during this process to ensure an appropriate subtraction was performed. It was noted that scaling did not change the main features in the final map, however it did change the contrast. The scaling was performed so as to provide the best contrast in the final map. Each of the smoothed, normalised histograms are illustrated in Figure \ref{fig:hist}, as well as the final density map. Note that the inter-chip gaps between CCDs are not visible in the final smoothed, subtracted density map. The position of the two bright stars in the field are also visible as a white gap in the map (this part of the map is assigned a ``NaN" value and treated as such in the analysis).

\begin{figure*}
\centering
\includegraphics[width=12cm]{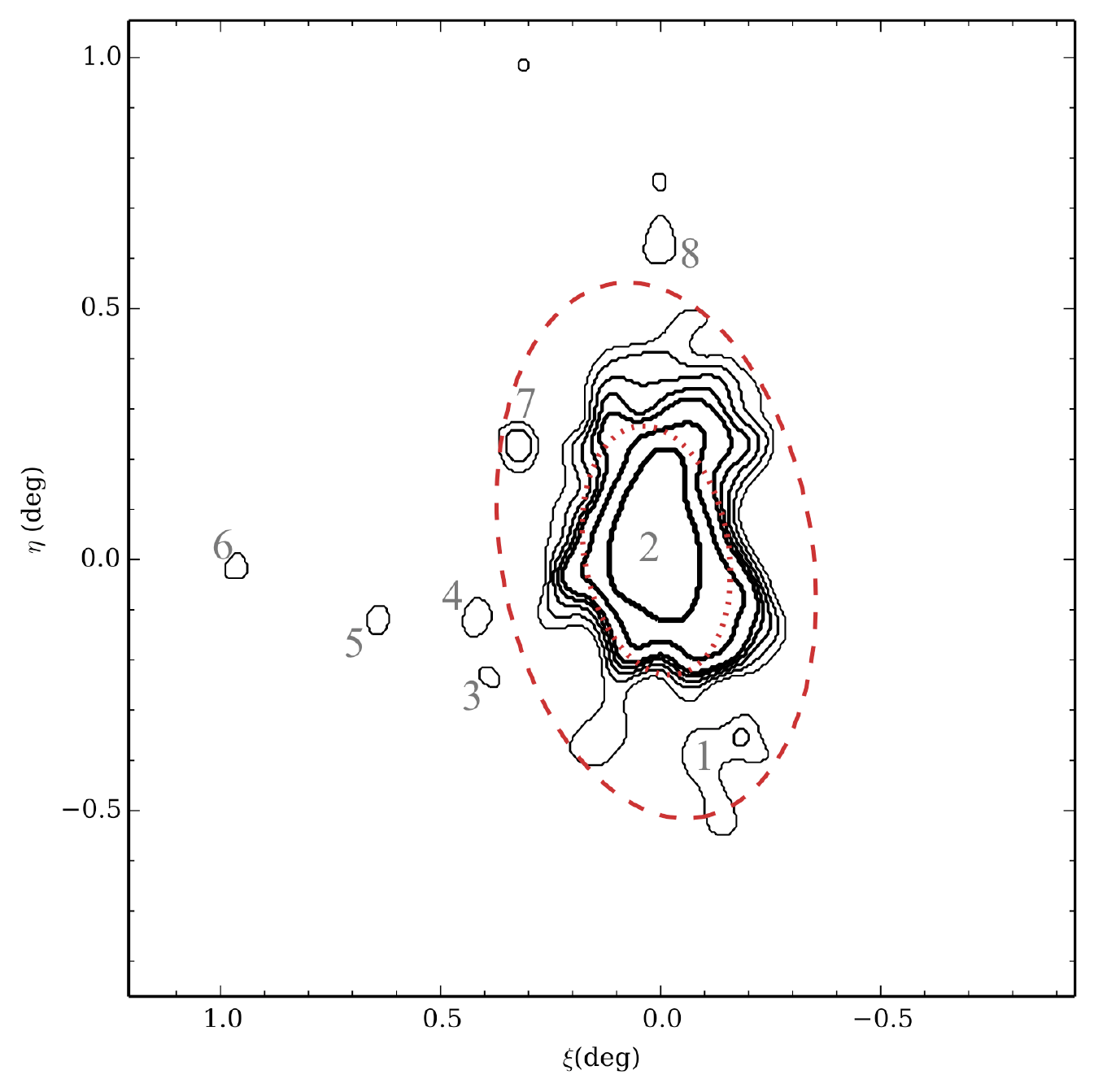}
\caption{Contour plot showing the foreground subtracted spatial distribution of Bo\"{o}tes\,I. Contours (black solid lines) represent 3$\sigma_t$, 4$\sigma_t$, 5$\sigma_t$, 6$\sigma_t$, 7$\sigma_t$, 10$\sigma_t$, 15$\sigma_t$, and 20$\sigma_t$ above the background, where $\sigma_t$ is a measure of the standard deviation in the background and the background is measured as all pixels outside three times the half-light radius. Note that line width increases with increased detection threshold. Grey numbers illustrate the labelling scheme adopted and discussed in Section \ref{sec:analysis}. The red dashed and dotted lines represent the tidal and core radii respectively.}
\label{fig:contours}
\end{figure*}

In order to identify structure in the map, a measure of the average background pixel value was determined by averaging over all pixels outside three times the half-light radius. Using the standard deviation of the background, $\sigma_t$, contours are defined in units of $\sigma_t$ above the mean. These contours are used as the basis for detection thresholds later in the substructure analysis, with the subscript $t$ serving as a reminder that this standard deviation defines the thresholds. Figure \ref{fig:contours} displays contours from the density map at 3$\sigma_t$, 4$\sigma_t$, 5$\sigma_t$, 6$\sigma_t$, 7$\sigma_t$, 10$\sigma_t$, 15$\sigma_t$, and 20$\sigma_t$. As a means of confirming the robustness of our initial star/galaxy separation, we compared the contours with the distribution of background galaxies downloaded from SDSS, and found no correlation.

\section{Structural Parameters}
\label{sec:parameters}
\begin{figure}
\centering
\includegraphics[width=\columnwidth]{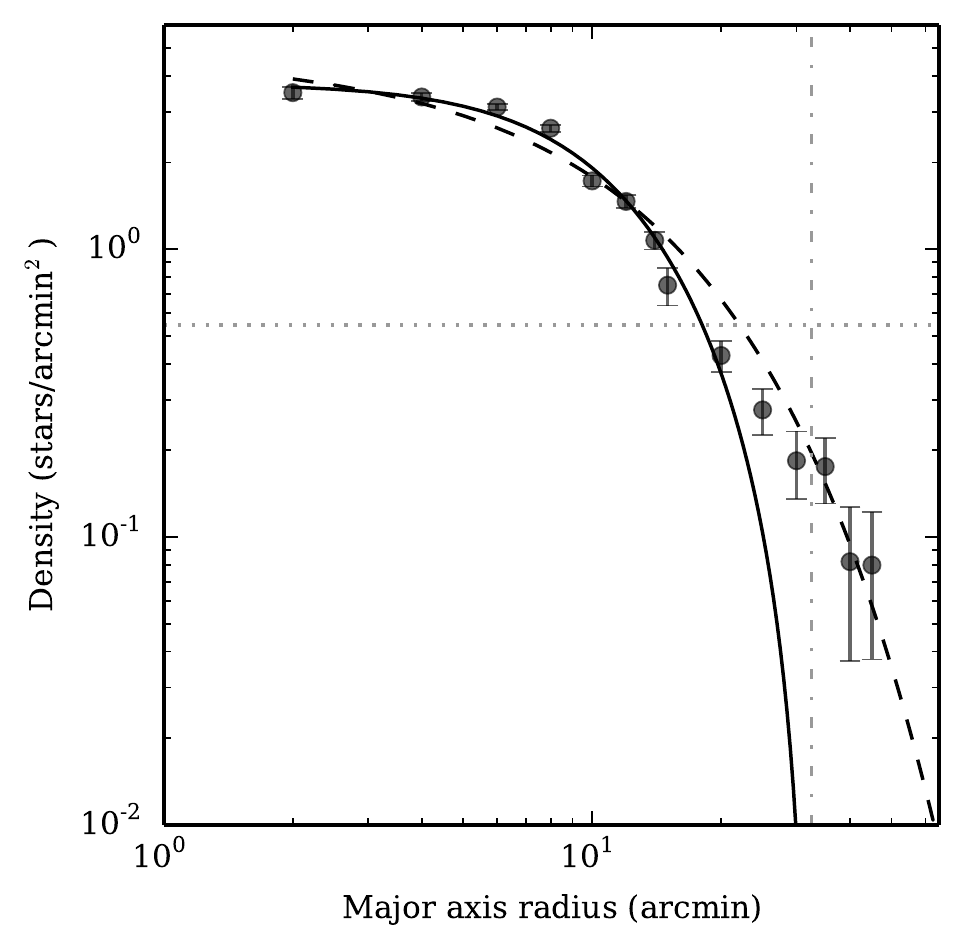}
\caption{Radial profile showing number density of stars vs. major axis distance from centre. The black solid line represents a King profile, while the black dashed line represents an exponential profile. Grey dotted and dash-dot lines represent the average background density and tidal radius respectively.}
\label{fig:radial}
\end{figure}

Prior to the substructure analysis, we determined the orientation angle, $\theta$, ellipticity, $\epsilon$ and centroid for Bo\"{o}tes\,I, and made a radial profile (shown in Figure \ref{fig:radial}). In a similar process to \citet{McConnachie:2006gl}, we used the contours from our density map in the determination of $\theta$, $\epsilon$ and the centroid. For each contour, at intervals of $1\sigma_t$ in the range between $3\sigma_t$ and $20\sigma_t$, a maximum likelihood bivariate normal fit was performed using the astroML\footnote{http://www.astroml.org/} \citep{2012cidu.conf...47V} machine learning package. The results were averaged across all contours, giving $\theta=13.4^\circ\pm8.6^\circ$, $\epsilon=0.33\pm0.06$ and a centroid offset in degrees from the central R.A. and Dec of $0.010\pm0.006$, $0.018\pm0.008$. These values are comparable to those found in the literature \citep[$0.22<\epsilon<0.39$, $\theta=14^\circ\pm6^\circ$,][]{2006ApJ...647L.111B,2008ApJ...684.1075M,2012ApJ...744...96O}. The irregular shape of Bo\"{o}tes\,I contributes to the large uncertainty in the orientation angle, however basing the measurement on the contours provides a robust estimate of the uncertainty intrinsic to this dwarf. Note that the centroid is varies from the centre of the $15\sigma_t$ contour by 0.002 and 0.01 degrees in R.A. and Dec respectively.

Using the values determined for $\theta$, $\epsilon$ and the centroid, we created elliptical annuli about the centre of Bo\"{o}tes\,I, with widths of $1.5-5$ arcminutes increasing radially outward until the limiting edge of the field (45 arcmin along the major axis from the centre of Bo\"{o}tes\,I).  The star density of each annulus was recorded, and the background density subtracted (found by averaging over random samples of the field).  Although there are inter-chip gaps present between the CCDs, these are regular and systematic across the field. Thus it was deemed that they would not affect the overall outcome of the profile.  The resulting profile is shown in Figure \ref{fig:radial}.  Both a King profile \citep{1962AJ.....67..471K} and exponential profile were fit to the data.  The King profile provides a better fit to the central region ($<20$ arcmin) than the exponential profile.  However, there appears to be a break radius at this point, where the exponential profile provides a better fit.  According to the King profile, the tidal radius is $r_t = 32\farcm5\pm3\farcm4$, with a core radius of $r_c = 15\farcm1\pm1\farcm6$, placing the break radius inside the tidal radius.  This may indicate that Bo\"{o}tes\,I has underlying substructure. We will discuss the break radius further in Section \ref{sec:disc}.

\section{Structural Analysis}
\label{sec:analysis}
This section contains the identification and quantification of substructure associated with Bo\"{o}tes\,I, and provides the focus of this paper. The analysis process is comprised of several parts; the explicit identification of substructure is based on the density map developed in Section \ref{sec:mapping}, and followed by quantification of each individual over-density. We also include a brief analysis of the distribution of candidate BSS and BHB stars. Each part of the analysis is described in the following.

\subsection{Identification of Substructure}
Since this paper is most concerned with the outskirts of Bo\"{o}tes\,I, and the central region provides sufficient signal to dominate the statistical analysis, we exclude the region inside the $15\sigma_t$ contour and concentrate on the structure outside this region. The density map was then `segmented' using the Python package \textit{scipy.ndimage}\footnote{http://docs.scipy.org/doc/scipy/reference/ndimage.html}, and individual structures given a segment identification number \citep[see][]{Roderick:2015jj,submitted}. The segmentation process was performed at six different detection thresholds, where the thresholds are given by the 3$\sigma_t$, 4$\sigma_t$, 5$\sigma_t$, 6$\sigma_t$, 7$\sigma_t$, 10$\sigma_t$ contours. In each case, everything above the detection threshold was considered a potentially significant over-density. Additionally, we performed the segmentation process at a threshold of 2$\sigma_t$. Despite the fact that this threshold revealed a lot of `noisy' detections, the contour surrounding the main body of Bo\"{o}tes\,I was determined to be statistically significant during the course of analysis. While this detection threshold is not discussed further, the significant contour is presented in our summary.

There was a total of eight detections at the $3\sigma_t$ level, some of which were also detected at higher detection thresholds.  Where there are repeat detections, the naming scheme of the $3\sigma_t$ level is adopted. The detections and labelling scheme are illustrated in Figure \ref{fig:contours}.  Once formally identified, the stars associated with each individual over-density were extracted from the catalogue to test the level of significance. All stars in the over-density were extracted, regardless of their weight, to ensure unbiased testing.  A control sample of stars was also necessary to calibrate against.  This provided a comparison for our detections to assess their significance. We do not have a separate field for comparison, however the DECam field-of-view provides a large region of sky unoccupied by Bo\"{o}tes\,I. Since the stellar over-densities were identified as anything above the detection threshold, we define a suitable control region as any part of the sky below the detection threshold. This ensures a reasonable representation of the sky, making use of as much of the DECam field as possible without including any regions which may be associated to Bo\"{o}tes\,I.

\subsection{Quantification of Substructure}
We use the same method as \citet{Roderick:2015jj} to quantify the significance of our over-density detections. For the purpose of testing we consider a star to be in close proximity to the Bo\"{o}tes\,I isochrone if it is weighted $w\geq0.8$.

Using the previously defined control region, a sample of stars are drawn equal in number to the over-density being tested (for the purpose of testing each detection threshold independently of the others, we count only stars `between' thresholds, such that stars detected at higher thresholds are not counted in the lower thresholds).  In both the sample and the over-density, stars in close proximity to the isochrone ($w\geq0.8$) are counted and noted.  This is repeated for the control region 10,000 times, with the sample drawn randomly each time.  A frequency distribution is created from this, and a Gaussian function fit to provide a mean and standard deviation ($\sigma$).  The significance ($\zeta$) is then determined by comparing the number of stars in the over-density with $w\geq0.8$ ($N_{w\geq0.8}(OD)$) with the average value determined for the control sample ($N_{w\geq0.8}(CS)$).  The significance is defined as the separation between the two values in units of $\sigma$:

$$
\zeta = \frac{N_{w\geq0.8}(OD) - \langle N_{w\geq0.8}(CS)\rangle}{\sigma}
$$

\begin{figure}
\centering
\includegraphics[width=\columnwidth]{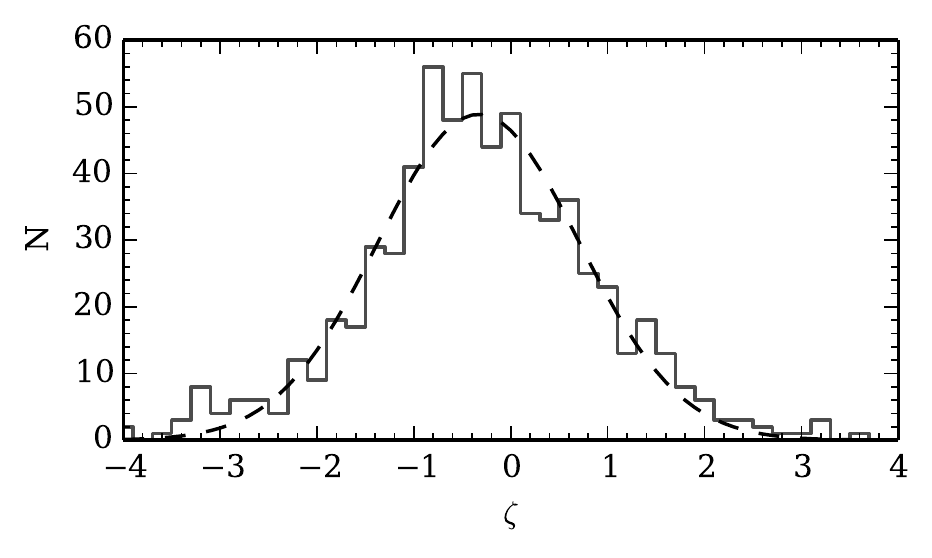}
\caption{The frequency distribution of $\zeta$ values resulting from a test of the null hypothesis.  This distribution shows a mean of $-0.33$ and standard deviation of $1.04$.}
\label{fig:null}
\end{figure}

\begin{figure*}
\centering
\includegraphics[width=\textwidth]{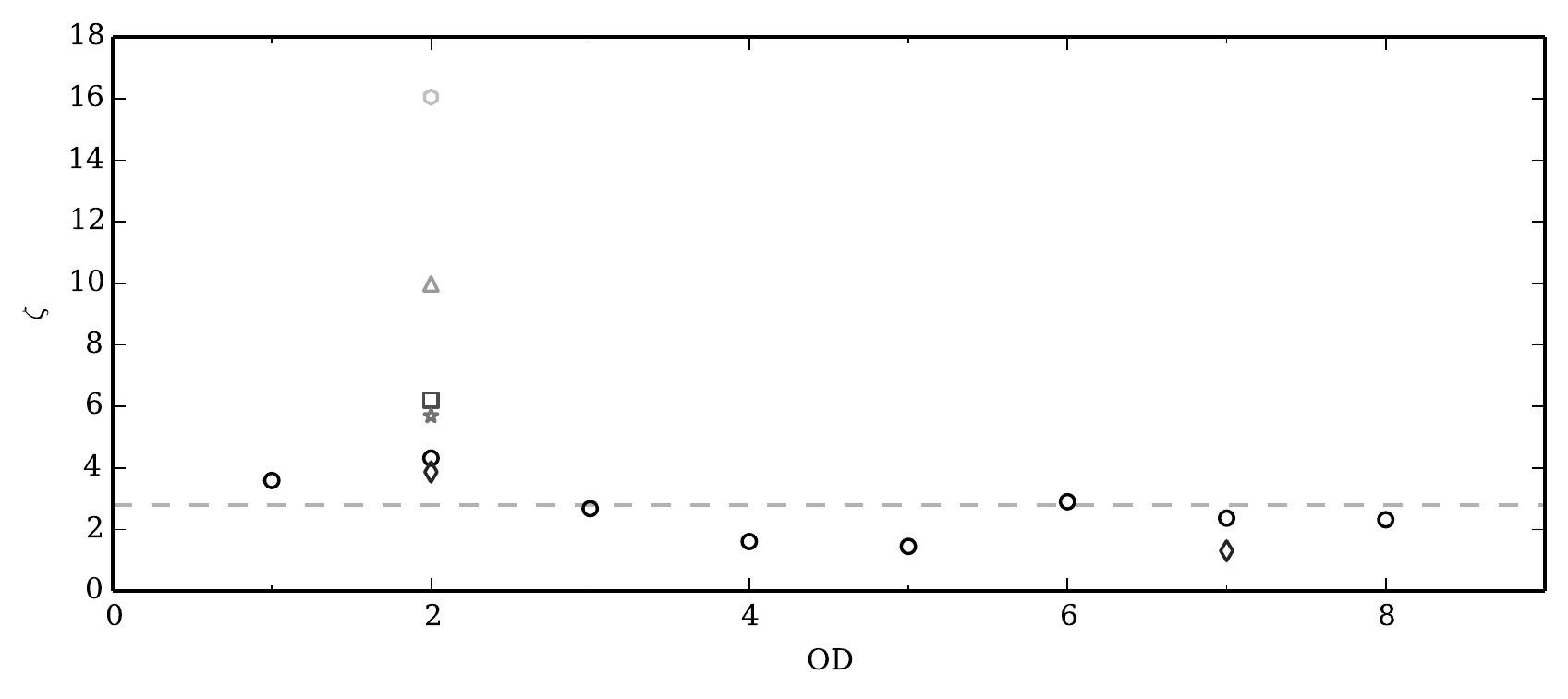}
\caption{Graphical representation of the significance of each detection (listed in Tables \ref{tab:zeta}). Circles, diamonds, squares, stars, triangles and hexagons represent thresholds $3\sigma_t$ to $10\sigma_t$ respectively. The grey dashed line represents $\zeta=2.79$}
\label{fig:results}
\end{figure*}

In order to assess the level of significance, we also performed a test of the null hypothesis.  We defined 550 random over-densities across the field, varying in size between 4 and 400 square arcminutes, but avoiding regions previously defined as candidate over-densities belonging to Bo\"{o}tes\,I.  We then performed the same significance testing using the control sample, with the expectation that the frequency distribution of $\zeta$ values should be centred on approximately zero (the frequency distribution is shown in Figure \ref{fig:null}). We fit a Gaussian function to the distribution, and found a mean of $-0.33$ and standard deviation of $1.04$. Based on this, we consider any detection with a significance value of three standard deviations or more above the mean to be statistically significant, providing a quantitative assessment value of $\zeta\geq2.79$. The results of the significance testing are shown in Figure \ref{fig:results}.  As can be seen from the figure, there are two main over-densities considered significant, with the main structure surrounding Bo\"{o}tes\,I detected at multiple levels.  The results of testing for each significant detection are listed in Table \ref{tab:zeta}.

\begin{table}
\begin{center}
\caption{Significance values for the statistically significant detections, where segment labelling corresponds to Figure \ref{fig:contours}. Each detection threshold is separated by a horizontal line, with the top rows corresponding to $3\sigma_t$, progressing down the table to $10\sigma_t$.}
\label{tab:zeta}
\begin{tabular}{ccrrrr}
\hline\hline
Segment && $N_{*}$ & $N_{w\geq0.8}(OD)$ & $\langle N_{w\geq0.8}(CS)\rangle$ & $\zeta$ \\
\cline{1-1}\cline{3-6}
OD 1 && 187 & 49 & 31 & 3.59 \\
OD 2 && 457 & 109 & 75 & 4.32 \\
OD 6 && 6 & 4 & 0 & 2.90 \\
\hline
OD 2 && 308 & 76 & 50 & 3.87 \\
\hline
OD 2 && 292 & 89 & 48 & 6.21 \\
\hline
OD 2 && 203 & 63 & 33 & 5.69 \\
\hline
OD 2 && 444 & 149 & 73 & 9.98 \\
\hline
OD 2 && 777 & 290 & 128 & 16.05 \\
\hline \hline
\end{tabular}
\end{center}
\end{table}

\begin{figure*}
\centering
\includegraphics[width=\textwidth]{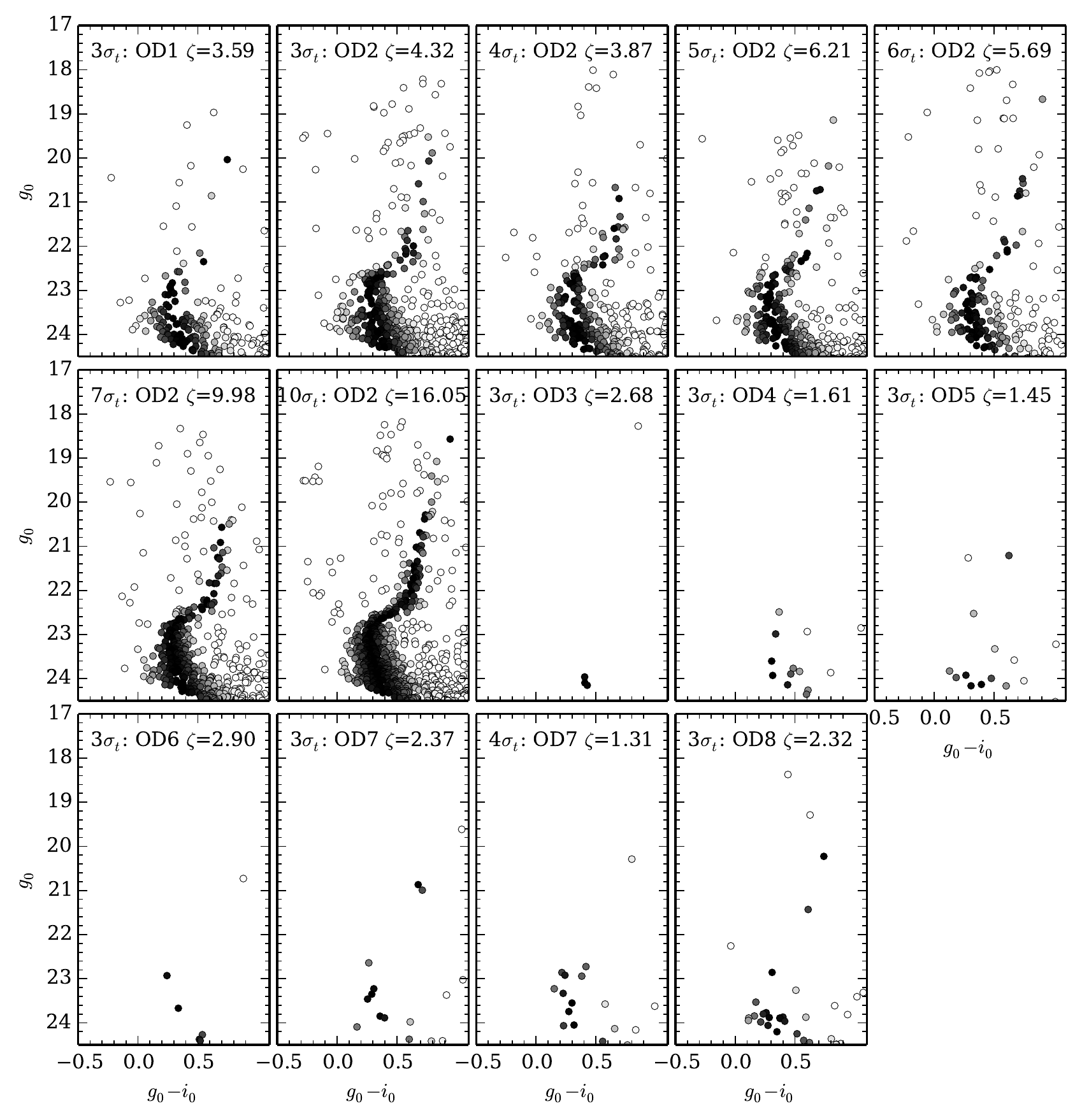}
\caption{Colour-magnitude diagrams of all detections. Points are coloured according to weight, with black representing 1.0, and white representing 0.0. Features of the Bo\"{o}tes\,I fiducial are clearly present.}
\label{fig:cmd}
\end{figure*}

To take this one step further, we made a visual inspection of the CMDs for each of the over-densities detected at each threshold (Figure \ref{fig:cmd}). The Bo\"{o}tes\,I fiducial is clearly present in OD 2 at each detection threshold, keeping in mind that these diagrams do not contain stars from the very centre of the dwarf ($>15\sigma_t$). The marginally significant OD 6 contains only 6 stars, making its significance unclear. However, the almost significant OD 8 forms part of the 2$\sigma_t$ structure which is statistically significant. This suggests it is plausible that the Bo\"{o}tes\,I substructure does extend as far north as OD 8.

As a final check for our over-densities, we made a comparison between their spatial distribution and the distribution of background galaxies in the field \citep[using the galaxy catalogue from][]{2012A&A...540A.106T}. No correlation was found, supporting the stellar nature of the over-densities.

An electronic supplement is available containing the coordinates of all stars in the detections determined to have significant $\zeta$ values.

\subsection{BSS and BHB Populations}
\begin{figure}
\centering
\includegraphics[width=\columnwidth]{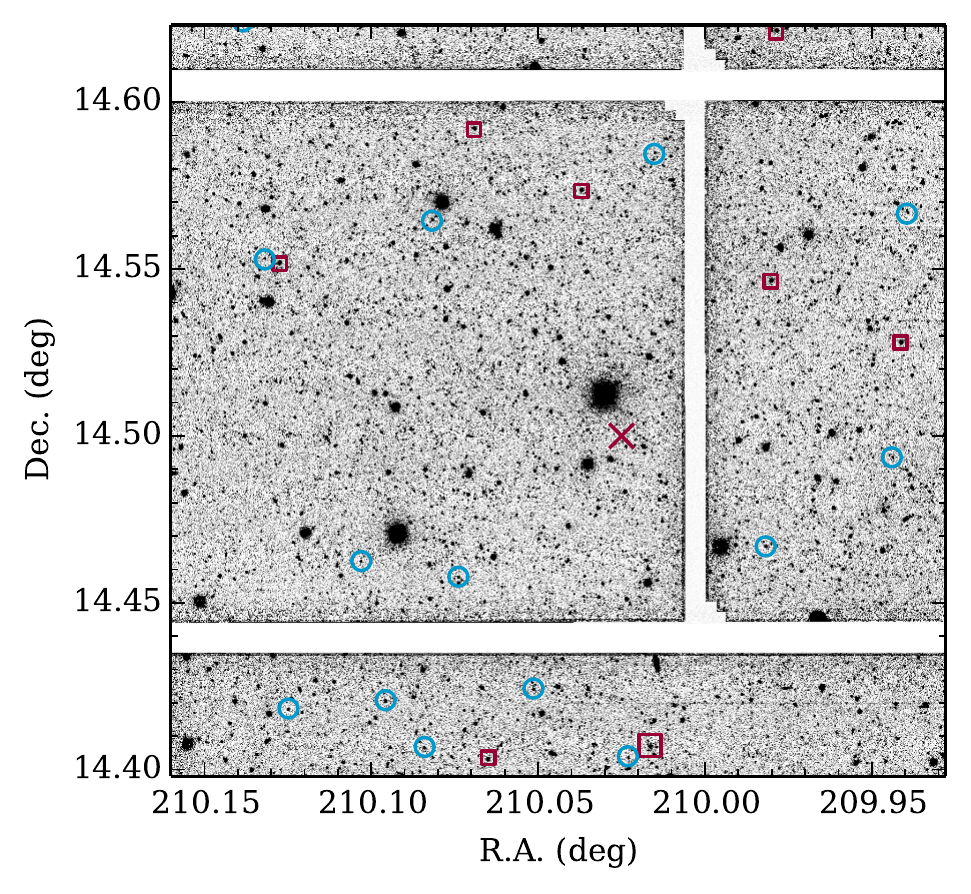}
\caption{Image in $g$-band of the central $12\arcmin\times12\arcmin$ region of Bo\"{o}tes\,I (corresponding to the dashed box in Figure \ref{fig:summary}). The nominal centre of Bo\"{o}tes\,I is marked with a cross. Note the absence of BHB (red squares) and BSS (blue circles) populations at the centre. Significant crowding or CCD inter-chip gaps do not appear to be a factor in the distribution of these populations. Note that the BHB stars are split into two groups by colour, where large squares represent $g-i <-0.22$ and small squares represent $g-i \geq -0.22$.}
\label{fig:stamp}
\end{figure}

\begin{figure}
\centering
\includegraphics[width=\columnwidth]{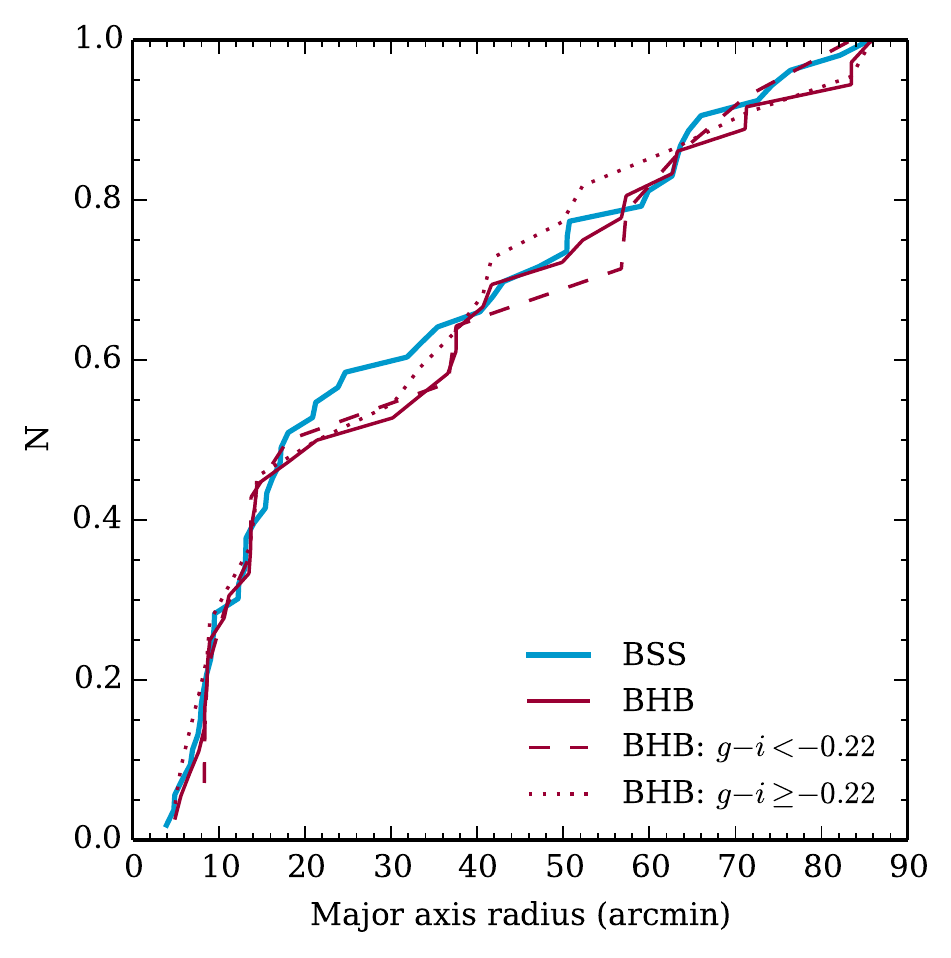}
\caption{Cumulative distribution functions for the BHB star and BSS populations.  The BHB star population has been further split into two groups at $g-i = -0.22$ to investigate the absence of stars with $g-i < -0.22$ in the centre of Bo\"{o}tes\,I.}
\label{fig:cumulative}
\end{figure}

As one last step to the analysis, we also made a brief investigation into the radial distribution of candidate BSS and BHB stars. It was noted that there appeared to be an absence of these populations from the very core of Bo\"{o}tes\,I. Whilst the absence of these stars appears to coincide with the central stellar density of Bo\"{o}tes\,I (see Figure \ref{fig:hist}), crowding or inter-chip gaps between CCDs does not appear to be a factor (see Figure \ref{fig:stamp}). The distance of each star from the centre of Bo\"{o}tes\,I was calculated in terms of the major axis distance, and a cumulative distribution function plot for each population (see Figure \ref{fig:cumulative}).  A Kolmogorov-Smirnoff (K-S) test was performed to test the likelihood of the two populations being drawn from the same distribution.  With a calculated $p$-value of 1.0, this test shows that to be highly likely. During the course of the analysis, it was also noticed that there appeared to be a difference in the radial distribution of bluer BHB stars. The difference was most notable at $g-i$ = -0.22, and can be observed in Figure \ref{fig:stamp}; inside the $12\arcmin\times12\arcmin$ frame of the image, there is only one bluer star compared to seven redder stars. To investigate this further, the BHB star population was split in two at this point and a cumulative distribution function and K-S test performed (see dashed and dotted lines in Figure \ref{fig:cumulative}. This test also yielded a $p$-value of 1.0, demonstrating a high likelihood that the two samples do belong to the same population. However, with so few stars, it is difficult to analyse the absence of various stellar populations in the core of Bo\"{o}tes\,I further.

\section{Discussion}
\label{sec:disc}
\begin{figure*}
\centering
\includegraphics[width=\textwidth]{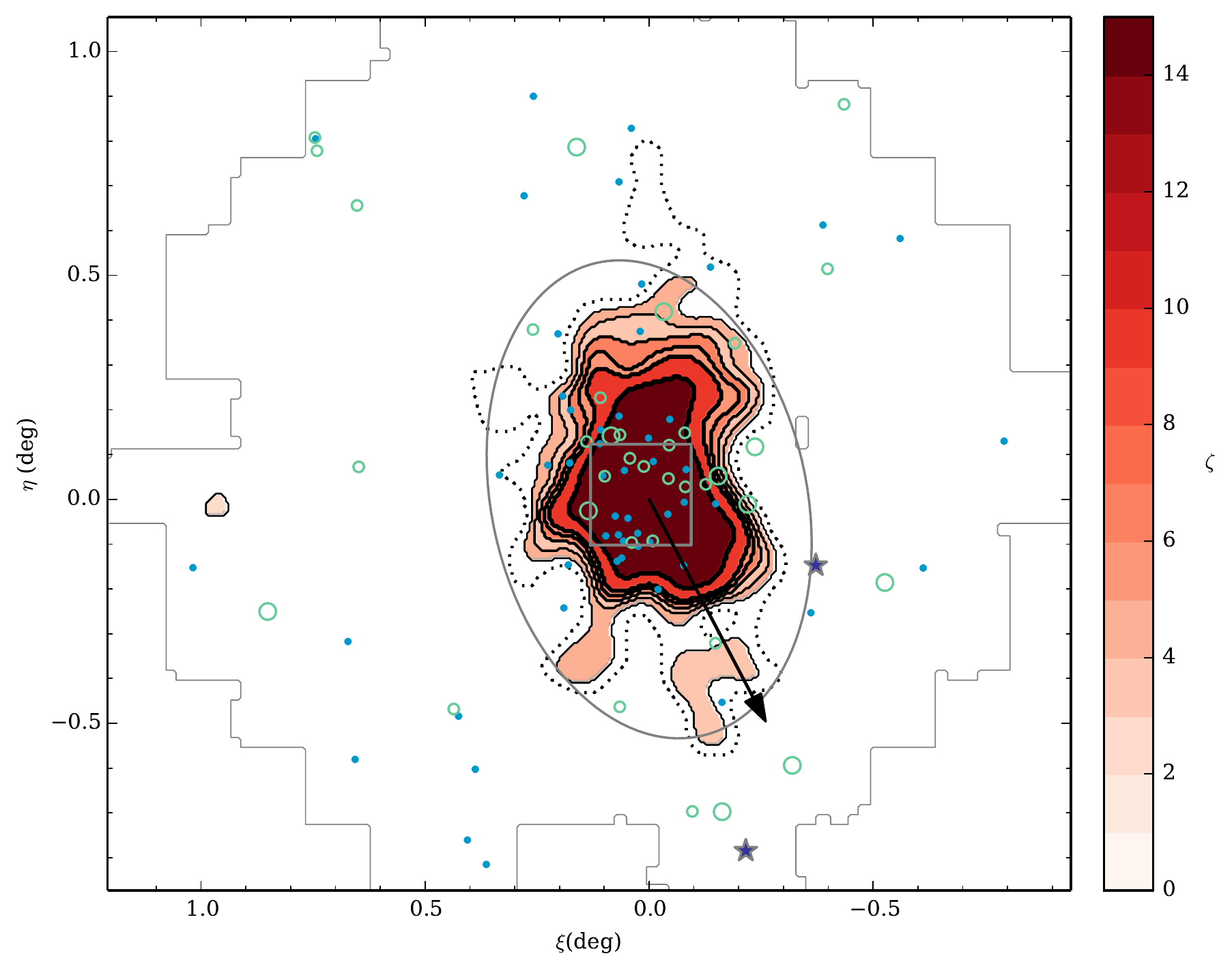}
\caption{Summary figure of the results of our substructure detection and significance testing across all detection thresholds. Contours show an increasing line-width with an increase in detection threshold between 3$\sigma_t$ and 10$\sigma_t$. A dashed contour with $\zeta=6.19$ for the $2\sigma_t$ threshold is shown without a fill colour. All other detection thresholds are over-laid on top of each other, and colour coded according to the $\zeta$ value. Open, green circles represent the Blue Horizontal Branch stars (outlined in red in Figure \ref{fig:mask}), while the blue, filled circles represent Blue Straggler stars (outlined in blue in Figure \ref{fig:mask}). Note that the Blue Horizontal Branch stars are split into two groups by colour, where large points represent $g-i <-0.22$ and small points represent $g-i \geq -0.22$. The tidal radius is represented by a grey ellipse. The box outlined by a dashed line at the centre of the figure corresponds to the region displayed in Figure \ref{fig:stamp}.The black arrow represents the model orbital path favoured by \citet{Fellhauer:2008dt}. The two purple-filed stars represent Boo-980 and Boo-1137, identified as member stars by \citet{2008ApJ...689L.113N} using radial velocity measurements.}
\label{fig:summary}
\end{figure*}

The elongation and irregular shape of Bo\"{o}tes\,I \citep{2006ApJ...647L.111B,Fellhauer:2008dt,2012ApJ...744...96O} suggests the possibility that this dwarf galaxy is in the process of being tidally disrupted. The over-densities we have detected show remarkable correlation with the contours shown in Figure 1 of \citet{Fellhauer:2008dt}, and the overall shape appears similar to that of \citet{2012ApJ...744...96O}.  Figure \ref{fig:summary} shows a summary of our analysis. Those over-densities determined to be statistically significant ($\zeta\geq2.79$) during the course of our analysis are shown at each detection threshold, colour coded according to their $\zeta$ value at a given threshold. Stars inside the central region ($>15\sigma_t$) were omitted from the analysis in order to reduce the effect of the dominant central region and reveal the underlying substructure.  The dashed contour represents a significant detection ($\zeta=6.19$) at the 2$\sigma_t$ threshold. This contour connects the main body of Bo\"{o}tes\,I with OD 1, and extends far enough north to encompass OD 8, an `almost' significant detection.

The detected substructure appears more extended towards the north of Bo\"{o}tes\,I, relative to the central coordinates.  The shape of the substructure surrounding Bo\"{o}tes\,I suggests the possibility of the beginning of ``S-shaped" tidal tails, also suggested by \citet{Fellhauer:2008dt}. However, the elongation and structure seen are not as extreme as that seen in Hercules \citep{Roderick:2015jj}. Also unlike Hercules, the BHB population (and additionally BSS) does not appear aligned in any particular direction.  These populations appear more uniformly dispersed like those seen in Sextans \citep{submitted}. There is a separate over-density (OD 1) toward the south-west of the centre, which is also in approximately the same direction as one of the model orbital paths suggested by \citet{Fellhauer:2008dt}. Brief investigation of the spatial distribution of BHB and BSS populations suggests a high likelihood that they are drawn from the same underlying population, consistent with the idea that Bo\"{o}tes\,I had a single period of star formation \citep{2006ApJ...647L.111B,2012ApJ...744...96O}. 

Figure \ref{fig:summary} also shows two stars (Boo-980, Boo-1137) identified as member stars by \citet{2008ApJ...689L.113N} using radial velocity measurements. Both stars are extremely metal poor and lie at large distances of 2.0 and 3.9 half-light radii from the centre of Bo\"{o}tes\,I. Boo-1137 has been the subject of detailed spectroscopic follow up \citep{2010ApJ...711..350N}. Both stars have been noted to show radial velocities in remarkable agreement with the mean velocity of the Bo\"{o}tes\,I system \citep{Koposov:2011kx}. While these stars do not directly coincide with our detections, they do align with the direction of extension of our substructure.

Based on a King model fit to our radial profile, we determined a tidal radius for Bo\"{o}tes\,I of $r_t = 32\farcm5\pm3\farcm4$ (shown as a grey ellipse in Figure \ref{fig:summary}). The most extended over-densities reach beyond the tidal radius, lending weight to the scenario in which Bo\"{o}tes\,I is being tidally disrupted. The radial profile also shows a break radius (at a major axis distance of approximately $20\arcmin$), indicating the presence of extra-tidal stars. The appearance of such a break radius is somewhat controversial in the MW dwarf spheroidals; despite the lack of an appearance in Draco and Sextans \citep[][]{2001AJ....122.2538O,2007MNRAS.375..831S,submitted}, the existence of a break radius in Carina and Sculptor is somewhat ambiguous \citep{2000AJ....120.2550M,2005AJ....130.2677M,2005AJ....130.1065C,2006AJ....131..375W,2014MNRAS.444.3139M}. Similar profiles have also been generated by \citet{2012ApJ...744...96O} for Canes Venatici\,I, Bo\"{o}tes\,I, Canes Venatici\,II, and Leo\,IV. Although they mention an excess in stellar density in the outskirts of Canes Venatici\,II, and Leo\,IV, they do not see such an occurrence in Bo\"{o}tes\,I.  This may in part be due to the fact that their radial profile does not extend beyond $30\arcmin$, where the excess in stellar density becomes more prominent.

A number of orbital scenarios have been put forward for Bo\"{o}tes\,I in an attempt to identify its most likely progenitor \citep{Fellhauer:2008dt}. The two main scenarios they suggest pose that either the progenitor was a purely baryonic star cluster being tidally disrupted, or that it was a dark matter dominated, cosmologically motivated system. The possibility was also considered that the extended stellar substructure surrounding Bo\"{o}tes\,I was the result of sparse stellar photometry, and therefore an observational artefact. The result of the analysis by \citet{Fellhauer:2008dt} ruled out the scenario of the baryonic star cluster because it could not reproduce the velocity dispersion observed.  However, this model reproduced the observed radial surface density profile quite well, as opposed to the dark matter dominated models, which did reproduce the velocity dispersion but not the radial surface density profile. Given the revised observed velocity dispersion of $\sigma=2.4^{+0.9}_{-0.5}$ km s$^{-1}$\citep{Koposov:2011kx} (note this is corresponds to the dominant kinematic component, which is much closer to that provided by the baryonic model of \citet{Fellhauer:2008dt}, the best estimate for a single component dispersion is 4.6 km s$^{-1}$), and our own analysis demonstrating the statistical significance of the extended substructure surrounding Bo\"{o}tes\,I, this suggests that the possible orbital scenarios for Bo\"{o}tes\,I should be reconsidered. Our analysis clearly demonstrates that the substructure seen around Bo\"{o}tes\,I is not an observational artefact, and therefore cannot be ignored in the modelling of orbital scenarios.

The orbital direction favoured by \citet{Fellhauer:2008dt} is shown in Figure \ref{fig:summary} as a black arrow. This orbit would place Bo\"{o}tes\,I relatively close to an apo-galacticon of $76\,$kpc, with a peri-galacticon of $37\,$kpc. Given the lack of abundant free-floating debris compared to Hercules \citep{Roderick:2015jj}, Bo\"{o}tes\,I is unlikely to follow as extreme an orbit as the model proposed for Hercules, particularly given it has an estimated infall time of $7-10\,$Gyr compared to the $2-8\,$Gyr estimated for Hercules \citep{2012MNRAS.425..231R}. 

\section{Summary}
\label{sec:sum}
We have demonstrated quantitatively that the Bo\"{o}tes\,I dwarf spheroidal galaxy shows significant, extended stellar substructure. An analysis of the radial profile has also revealed the presence of extra-tidal stars, providing strong evidence to suggest that Bo\"{o}tes\,I is experiencing tidal disruption. In conjunction with the revised velocity dispersion measurements from \citet{Koposov:2011kx}, this suggests that Bo\"{o}tes\,I may not possess the high mass-to-light ratio previously thought, and suggests a review of previous orbital models and possible progenitors may be necessary.

\section*{Acknowledgements}

The authors wish to thank the referee for their comments on the original manuscript. TAR acknowledges financial support through an Australian Postgraduate Award. ADM acknowledges the support of the Australian Research Council through Discovery Projects DP1093431, DP120101237, and DP150103294. HJ and GDC acknowledge the support of the Australian Research Council through Discovery Projects DP120100475 and DP150100862. This project used data obtained with the Dark Energy Camera (DECam), which was constructed by the Dark Energy Survey (DES) collaborating institutions: Argonne National Lab, University of California Santa Cruz, University of Cambridge, Centro de Investigaciones Energeticas, Medioambientales y Tecnologicas-Madrid, University of Chicago, University College London, DES-Brazil consortium, University of Edinburgh, ETH-Zurich, Fermi National Accelerator Laboratory, University of Illinois at Urbana-Champaign, Institut de Ciencies de lÕEspai, Institut de Fisica dÕAltes Energies, Lawrence Berkeley National Lab, Ludwig-Maximilians Universitat, University of Michigan, National Optical Astronomy Observatory, University of Nottingham, Ohio State University, University of Pennsylvania, University of Portsmouth, SLAC National Lab, Stanford University, University of Sussex, and Texas A\&M University. Funding for DES, including DECam, has been provided by the U.S. Department of Energy, National Science Foundation, Ministry of Education and Science (Spain), Science and Technology Facilities Council (UK), Higher Education Funding Council (England), National Center for Supercomputing Applications, Kavli Institute for Cosmological Physics, Financiadora de Estudos e Projetos, Fundao Carlos Chagas Filho de Amparo a Pesquisa, Conselho Nacional de Desenvolvimento Cientfico e Tecnolgico and the Ministrio da Cincia e Tecnologia (Brazil), the German Research Foundation-sponsored cluster of excellence ÒOrigin and Structure of the UniverseÓ and the DES collaborating institutions.




\bibliographystyle{mnras}
\bibliography{BootesI} 

\bsp	
\label{lastpage}
\end{document}